\documentclass[aps,prb,twocolumn,amsmath,amssymb,preprintnumbers,floatfix,citeautoscript]{revtex4}

\usepackage[utf8]{inputenc}
\usepackage{txfonts}    
\usepackage{txgreeks}
\usepackage{microtype}  
\usepackage{eucal}  
\usepackage{bm} 

\usepackage[colorlinks,allcolors=blue]{hyperref}
\usepackage[capitalize]{cleveref}

\usepackage{enumerate}
\usepackage{xcolor}
\usepackage{braket}

\usepackage{graphicx}
\usepackage{float}

\usepackage{mdframed}    

\newcommand{\bcomm}[2]{\big[{#1}\,,\,{#2}\big]}                 

\renewcommand{\d}[1]{\mathrm{d}{#1}}                            

\renewcommand{\v}[1]{\boldsymbol{#1}}                           

\newcommand{\up}{\uparrow}                                      
\newcommand{\dn}{\downarrow}                                    

\newcommand{\refcite}[1]{Ref.\phantom{-}\onlinecite{#1}}        

\definecolor{DarkRed}{rgb}{0.65,0,0}
\definecolor{DarkBlue}{rgb}{0,0,0.65}
\definecolor{DarkGreen}{rgb}{0,0.65,0}

\begin{document}
\title{Conservation of spin supercurrents in superconductors}
\author{Jabir Ali Ouassou}
\author{Sol H. Jacobsen}
\author{Jacob Linder}
\affiliation{{QuSpin Center of Excellence,} Department of Physics, Norwegian University of Science and Technology, N-7491 Trondheim, Norway}
\date{\today}

\begin{abstract}
  \noindent
  We demonstrate that spin supercurrents are conserved upon transmission through a conventional superconductor, even in the presence of spin-dependent scattering by impurities with magnetic moments or spin-orbit coupling.
  {This is fundamentally different from conventional spin currents, which decay in the presence of such scattering, and this has important implications for the usage of superconducting materials in spintronic hybrid structures.}
\end{abstract}

\maketitle


\noindent
\section{Introduction}
Superconducting spintronics marries the dissipationless currents {of superconductors} with nanoscopic hardware in which both charge and spin degrees of freedom {are} manipulated \cite{LinderRobinson2015,EschrigRev2011,EschrigRev2015,BlamireRobinson2014}.
Recent experiments have {demonstrated that such devices offer significant improvements over their non-superconducting counterparts, showcasing e.g.} enhanced spin lifetimes~\cite{Yang2010}, relaxation lengths~\cite{Quay2013}, spin Hall effects~\cite{Wakamura2015}, and infinite magnetoresistance~\cite{Li2013}.

Magnetic and spin-orbit impurities {usually act as antagonists to spintronics as they} cause a rapid spatial decay of \emph{conventional} spin currents, which remain polarized only up to the spin relaxation length~\cite{BassPratt2007,OtaniKimura2011}.
{Realistic superconductors also contain magnetic and spin-orbit impurities that randomize the electron spins. Taken in combination with the fact that Cooper pairs in the bulk of a conventional superconductor are spinless, one might expect such impurities to also cause a decay of spin \emph{super}currents in a superconductor.}
However, it has {recently} been shown that spin supercurrents do not decay in homogeneous magnets with spin-dependent scattering~\cite{JKL2016}.
Here, we find the surprising result that the spin supercurrent is also conserved in superconductors, and we analyze the underlying physical mechanism.
The result has significant implications for the use of superconductors in spintronic architectures, the experimental implementation of which is just starting to flourish~\cite{LinderRobinson2015,EschrigRev2011,EschrigRev2015,BlamireRobinson2014,Yang2010,Quay2013,Wakamura2015,Li2013}.

\section{Physical system}
\begin{figure}[tb]
  \centering
  \includegraphics[width=0.45\textwidth]{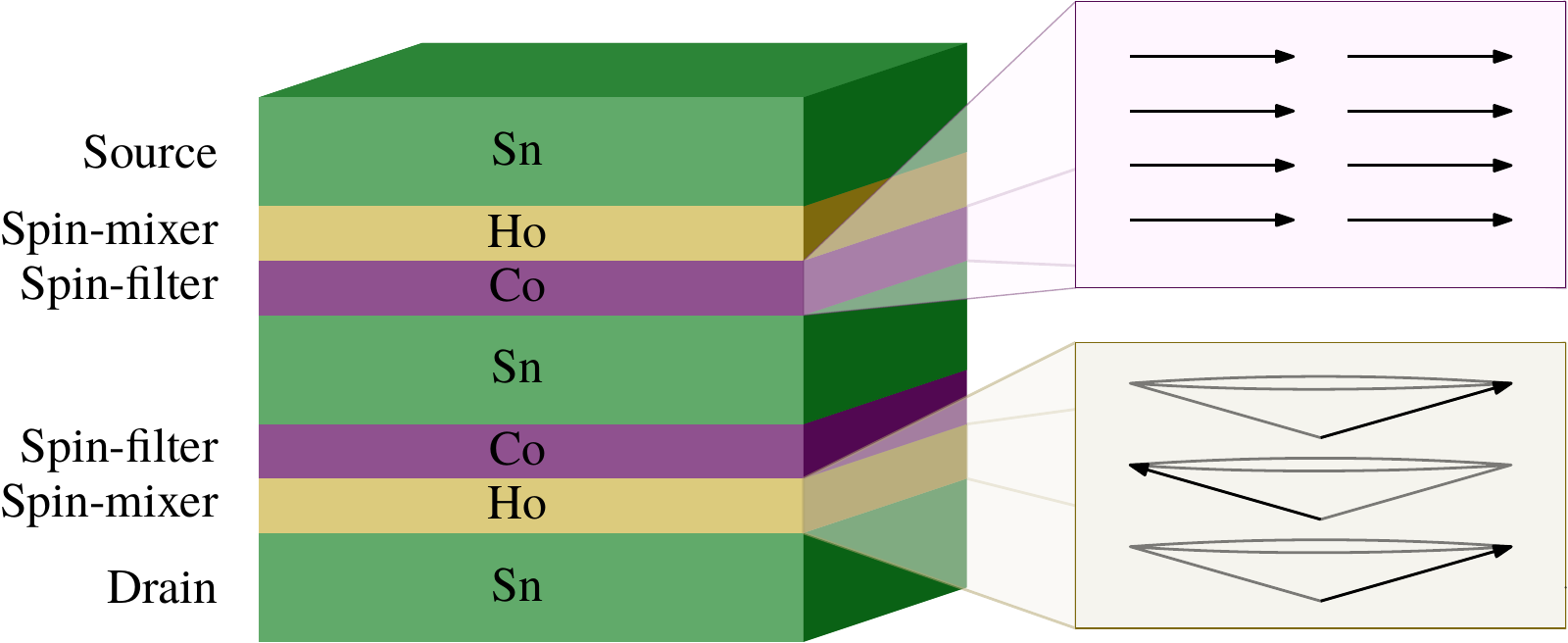}
  \caption
  {
    The thin-film stack considered in this paper.
    The insets on the right show the magnetization textures in the Co and Ho layers.%
  }%
  \label{fig:structure}
  \vspace{-3ex}
\end{figure}

We consider {theoretically} the spin supercurrents in a Sn/Ho/Co/Sn/Co/Ho/Sn multilayer stack, as shown in \cref{fig:structure}.
This structure is inspired by the Nb/Ho/Co/Ho/Nb system investigated experimentally in \refcite{Robinson2010};
however, we insert a superconductor within the central magnetic layer to investigate how spin currents behave in superconductors. 
{A related model was considered in \refcite{Alidoust2014}, which however focused on the behaviour of the charge and not spin supercurrents. In contrast to their model, we also treat the central superconductor selfconsistently in order to ensure charge conservation.}

Tin (Sn) is a conventional superconductor with a zero-temperature gap~$\Delta_0 \approx 1.15~$meV, coherence length $\xi \approx 30$~nm{, and density of states $N_0 \approx 66$~eV$^{-1}$~nm$^{-3}$ at the Fermi level}~\cite{SnDos}.
{Note that the value for the density of states corresponds to white tin at room temperature, but we will use it as an order-of-magnitude estimate also at cryogenic temperatures.}
The source and drain are bulk {superconductors}  with gaps $\Delta_0 e^{\pm i\varphi/2}$, where $\varphi$ is their phase-difference, while the gap is determined selfconsistently in the central Sn~\cite{Jacobsen2015}.
Holmium (Ho) is a conical antiferromagnet below {$\sim$}20\,K~\cite{Koehler1966}, having a spatially rotating in-plane magnetization with a constant out-of-plane component.
We therefore model the exchange field {in the Ho layer as}
\begin{equation}
  \v{h} = 5\Delta_0 \left[\sin(\gamma) \cos(z/{\zeta}),\, \sin(\gamma) \sin(z/{\zeta}),\, \cos(\gamma) \right],
\end{equation}
where the cone angle $\gamma = 0.45\pi$ \cite{Koehler1966}, and the spiral length ${{\zeta} = 3.34}$~nm \cite{Chiodi2013}.
We set the Ho layer lengths to ${1.5{\zeta} = 5}$~nm, since the spin current in the central layer is maximal for nonintegral numbers of Ho spirals~\cite{Robinson2010}.
Lastly, we model the relatively strong exchange fields of both Cobalt (Co) layers as $\v{h} = 50\Delta_0 \v{e}_y$, assuming that their magnetizations are homogeneous, parallel, and in-plane.
The lengths of each Co layer was set to $0.1\xi = 3$~nm.
Note however that the results are qualitatively unaffected by {these} specific parameter choices.

The outer Sn layers in \cref{fig:structure} are superconducting reservoirs, and act as sources for singlet Cooper pairs {$\ket{\up\dn}-\ket{\dn\up}$}.
These singlet pairs leak into the neighbouring Ho layers, where the inhomogeneous magnetic order converts them into a mixture of all possible triplet pairs: $\ket{\up\dn}+\ket{\dn\up}$, $\ket{\up\up}$, and $\ket{\dn\dn}$, {with spin projections} measured relative to the Co magnetization. 
The $\ket{\up\dn}+\ket{\dn\up}$ contributions rapidly decay inside Co, because the constituent electrons belong to different spin bands.
Furthermore, the {interfacial spin polarization} leads to a slight increase in the number of $\ket{\up\up}$ pairs compared to~$\ket{\dn\dn}$ pairs.
Thus, the central Sn layer obtains a proximity-induced repository of triplet pairs, which is dominated by the $\ket{\up\up}$ component, but also {contains} traces of other triplet components.
In addition, this Sn layer has its own condensate of singlet pairs, which may be converted into $\ket{\up\dn}+\ket{\dn\up}$ triplets at the Co interfaces.
This leads to a complex mixture of Cooper pairs throughout the junction, which may partake in the transmission of a completely general spin supercurrent.
In the rest of the manuscript, we analyze the properties and behaviour of this current.%

\section{Theory}
For the numerical analysis, we employ the quasiclassical theory of superconductivity, taking the whole junction to be diffusive and in equilibrium.
Specifically, we solve simultaneously the Usadel diffusion equation in each layer \cite{Usadel1970,Rammer1986,Chandrasekhar2004,Belzig1999,Silaev2015},
\begin{equation}
  i\hbar D \partial_z (\hat{g} \partial_z \hat{g}) \\ = \bcomm{\hat{\Sigma}}{\hat{g}},
  \label{eq:usadel}
\end{equation}
and the gap equation in the central superconductor~\cite{Jacobsen2015}, 
\begin{equation}
  \Delta(z) = \frac{1}{2}N_0\lambda\! \int^{+\Theta}_{-\Theta} \!\!\d\epsilon\, f_s(z,\epsilon) \tanh(\epsilon/2T),
  \label{eq:gapequation}
\end{equation}
yielding selfconsistent results.
\Cref{eq:usadel} is solved for the ${4\times4}$ matrix $\hat{g}(z,\epsilon)$, which contains the spin-resolved normal and anomalous retarded propagators as functions of the quasiparticle energy~$\epsilon$ and position~$z$.
The other matrices are
  \begin{align}
    \hat{\Sigma}      &= \epsilon{\hat{\tau}_3} + \hat{\Delta} + \bm{h}\hat{\bm\sigma} + \alpha_{\text{sf\,}} \hat{\bm\sigma}\hat{g}\hat{\bm\sigma} + \alpha_\text{so\,}{\hat{\tau}_3}\hat{\bm\sigma}\hat{g}\hat{\bm\sigma}{\hat{\tau}_3} , \\
    {\hat{\tau}_3}        &= \text{diag}(+1,+1,-1,-1), \\
    \hat{\Delta}      &= \text{antidiag}(+\Delta,-\Delta,+\Delta\!{}^*,-\Delta\!{}^*), \\
    \hat{\bm\sigma}   &= \text{diag}(\bm{\sigma}, \bm{\sigma}^*), 
  \end{align}
where $\bm\sigma$ is the Pauli vector.
In addition, the equations involve the local superconducting gap $\Delta$, exchange field $\bm{h}$, spin-flip scattering rate $\alpha_\text{sf}$, spin-orbit scattering rate $\alpha_\text{so}$, normal-state density of states at the Fermi level $N_0$, BCS coupling constant~$\lambda$, diffusion coefficient $D = \Delta_0\xi^{2\!}/\hbar$, temperature $T$, Debye cutoff $\Theta=\Delta_0\cosh(1/N_0\lambda)$, and Planck's reduced constant~$\hbar$.
Inelastic scattering is approximated by a complex quasiparticle energy, i.e. $\epsilon \rightarrow \epsilon + 0.01\Delta_0i$.
Only the singlet component $f_s$ of the propagator matrix $\hat{g}$ enters \cref{eq:gapequation}.
For more details about the numerical solution of these equations, see \refcite{Ouassou2016a}.

We use Kupriyanov--Lukichev boundary conditions for Sn/Ho interfaces \cite{Kupriyanov1988}, and general spin-active boundary conditions for Ho/Co and Co/Sn interfaces \cite{Eschrig2015,Machon2013,Cottet2007,Cottet2009,Cottet2011}.
To model experimentally realistic interfaces with low to moderate transparency, we set the ratio of tunneling to bulk conductance to be $G_T/G_0 = 0.3$ at each interface.
We set $G_\varphi/G_T = 0.3$ at each Co interface, where $G_\varphi$ is the spin-mixing conductance.
The interfacial spin-polarization of Co was set to $P = 0.12$, based on estimates for the polarization of the conductivity \cite{Villamor2013}.

Once the equations above are solved, the charge current $J_\mathrm{e}$ and spin current $\bm{J}_{\mathrm{\!s}}$ are found via
\begin{align}
  \!\!J_\mathrm{e}(z)        &= 2J_\mathrm{e0}\!\!   \int_{-{\infty}}^{+{\infty}} \!\!\!\!\!\!\d\epsilon\, \mathrm{Re}\,\mathrm{Tr} \big[ {\hat{\tau}_3} \hat{g}(z,\epsilon)\, \partial_{z\,} \hat{g}(z,\epsilon) \big] \tanh(\epsilon/2T),                              \\
  \!\!\bm{J}_\mathrm{\!s}(z) &= 2J_\mathrm{s0}\!\! \int_{-{\infty}}^{+{\infty}} \!\!\!\!\!\!\d\epsilon\, \mathrm{Re}\,\mathrm{Tr} \big[ \hat{\bm{\sigma}} {\hat{\tau}_3} \hat{g}(z,\epsilon)\, \partial_{z\,} \hat{g}(z,\epsilon) \big] \tanh(\epsilon/2T), \label{eq:spincurrent}
\end{align}
where ${J_\mathrm{e0} = eN_0^{\vphantom{2}} \Delta_0^2 \xi^2 \!A/ 4\hbar L}$, $J_\mathrm{s0} = \hbar J_\mathrm{e0}/2e$, $e$ is the electron charge, $L$ is the length of the central superconductor, and $A$ is the cross-sectional area of the junction.
{Using the previously specified material parameters for Sn, and assuming a superconductor length $L \approx \xi$, the current density unit can be estimated as $J_\mathrm{e0}/A \approx 16$~MA/cm$^2$.}
The charge and spin currents are measured relative to {this unit} throughout the manuscript{, and are typically 1--4 orders of magnitude smaller}.

The spin current above may be decomposed into an \emph{exchange current} $J_\mathrm{s+}$ and \emph{polarization current} $J_\mathrm{s-}$:
\begin{equation}
  \bm{J}_\mathrm{\!s\pm}(\varphi) \equiv [\bm{J}_\mathrm{\!s}(+\varphi) \pm \bm{J}_\mathrm{\!s}(-\varphi)]/2.
\end{equation}
By construction, these currents $J_\mathrm{s\pm}$ are symmetric and antisymmetric under $\varphi\rightarrow-\varphi$, respectively.
The polarization current is just the spin-polarized component of the charge current, and vanishes at $\varphi=0$ as expected.
Note that a polarization current can only be obtained if one includes the interfacial polarization of Co in the model, to act as a spin filter by providing different tunneling amplitudes for $\left|\up\up\right\rangle$ and $\left|\dn\dn\right\rangle$ Cooper pairs, where the spin-directions are measured relative to the Co magnetization.
Otherwise, the electric current will be transported by an equal number of $\left|\up\up\right\rangle$ and $\left|\dn\dn\right\rangle$ pairs, resulting in no net spin transfer.
{On the other hand,} the exchange current mediates the exchange interaction between ferromagnetic layers~\cite{Chen2014}, and can be finite even for $\varphi=0$, i.e. without any charge current.
These quantities behave quite differently as functions of the superconducting phase-difference \cite{JKL2016}.
Herein, we focus on the magnitudes $J_\mathrm{s\pm} = |\bm{J}_\mathrm{s\pm}|$ of these spin currents, and not {their polarization directions}.

{
In Josephson junctions~\cite{Josephson1962}, the supercurrent depends on the phase-difference~$\varphi$ between the superconductors as well as the properties of the barrier~\cite{Golubov2004}.
Whereas it takes a $\sin(\varphi)$ form for normal tunneling barriers, it can take a $\sin(\varphi+\pi)$ form when the barrier is magnetic~\cite{Golubov2004,Buzdin2005,Ryazanov2001}, and an even more exotic $\sin(\varphi+\varphi_0)$ form in the presence of spin-orbit coupling~\cite{Szombati2016,Buzdin2008,Reynoso2008,Zazunov2009,Tanaka2009}.
Herein, we are interested in studying supercurrents \emph{inside} a superconductor, which is why focus on double-barrier junctions with a superconductor sandwiched in the middle.
Depending on the system parameters, the supercurrent can also take a more complex $\sin(\varphi/2)\, \text{sgn}[\cos(\varphi/2)]$ form in such junctions~\cite{OuassouSinphin,Zapata1996,Ishikawa2001,Yu1999,DeLuca2009,Beenakker2013}.
This $\sin(\varphi/2)$ behaviour can intuitively be understood if one thinks of the double-barrier junction as a concatenation of two regular Josephson junctions, where a total phase-difference~$\varphi$ over the double-barrier junction is distributed evenly between the subjunctions.
In this manuscript, we will briefly discuss the charge supercurrent for completeness, but the main focus will however be the \emph{spin supercurrents}.
}

\section{Analytical argument}
We will now prove analytically that the spin supercurrent is conserved in superconductors with spin-flip and spin-orbit scattering {within a selfconsistent Born approximation framework}, where conventional spin currents typically decay \cite{Morten2005,Morten2004}. 
First, let us rewrite the spin current in {\cref{eq:spincurrent}} as 
\begin{align}
  \bm{J}_{\!\mathrm{s}}(z) = 2J_{\mathrm{s}0}\!\! \int_{-{\infty}}^{+{\infty}} \!\!\!\!\!\!\d\epsilon\, \mathrm{Re}[\bm{j}_\mathrm{s}(z,\epsilon)]\, \mathrm{tanh}(\epsilon/2T),
\end{align}
where we define a spectral spin current
\begin{align}
  \label{eq:spincurrentdens}
  \bm{j}_{\mathrm{s}} \equiv \mathrm{Tr} \big[ \hat{\bm{\sigma}} {\hat{\tau}_3} \hat{g} \partial_{z} \hat{g} \big].
\end{align}
The easiest way to satisfy the conservation of spin current~$\bm{J}_{\!\mathrm{s}}$ would be if the spectral spin current~$\bm{j}_{\mathrm{s}}$ is conserved as well.
Mathematically, this can be checked by differentiating \cref{eq:spincurrentdens}, and investigating whether the resulting expression vanishes:
\begin{align}
  \partial_z\bm{j}_{\mathrm{s}} = \mathrm{Tr} \big[ \hat{\bm{\sigma}} {\hat{\tau}_3} \partial_z(\hat{g} \partial_{z} \hat{g}) \big] = 0.
\end{align}
Note that $\partial_z(\hat{g} \partial_z \hat{g})$ is just the left-hand side {of the Usadel equation [\cref{eq:usadel}], such that the condition simplifies to}
\begin{equation}
  \label{eq:supertrace}
  \mathrm{Tr}\Big\{ \hat{\bm\sigma}{\hat{\tau}_3} \bcomm{\hat{\Sigma}}{\hat{g}} \Big\} = 0 .
\end{equation}
This condition can be checked by inserting the most general form for $\hat{g}$, containing all allowed symmetry components of the normal and anomalous propagators, and explicitly evaluating the trace~\cite{JKL2016}.
The process is straight-forward but tedious, and therefore omitted here.
The result is that the condition holds in the presence of a superconducting gap~$\Delta$, spin-flip scattering~$\alpha_\mathrm{sf}$, and spin-orbit scattering~$\alpha_\mathrm{so}$.
This constitutes an analytical proof that spin currents are conserved in superconductors with spin-flip and spin-orbit scattering.


{However, t}he quasiclassical approach also allows one to go further, permitting a selfconsistent numerical investigation to expose the underlying physical mechanisms behind this conservation of the spin supercurrent in superconductors.
We focus on this {topic} in the remainder of the paper.

\begin{figure}[b]
    \centering
    \includegraphics{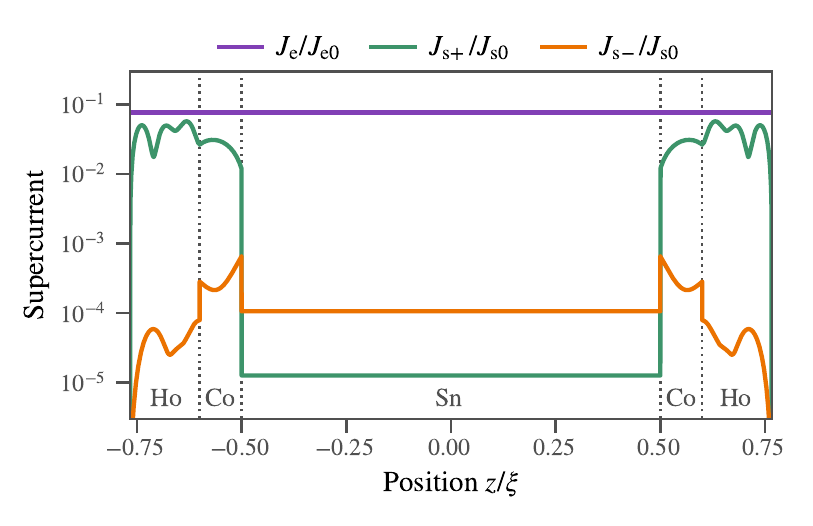}
    \caption
    {
      {Charge current~$J_\mathrm{\!e}$ and spin currents~$J_{\!\mathrm{s}\pm}$ as functions of position~$z$.}
      The dotted vertical lines indicate interfaces between materials in the junction.
      In this case, the phase-difference between the outer superconductors is $\varphi = \pi/2$, the central superconductor has length $L = \xi$, and the scattering rates are $\alpha_\text{sf} = \alpha_\text{so} = 0.01\Delta_0$.
      {Note that the charge current is conserved throughout the junction, while the spin current is only conserved inside the central superconductor.}
    }
    \label{fig:position}
\end{figure}
\section{Numerical results} 
\begin{figure}[b!]
    \centering
    \includegraphics{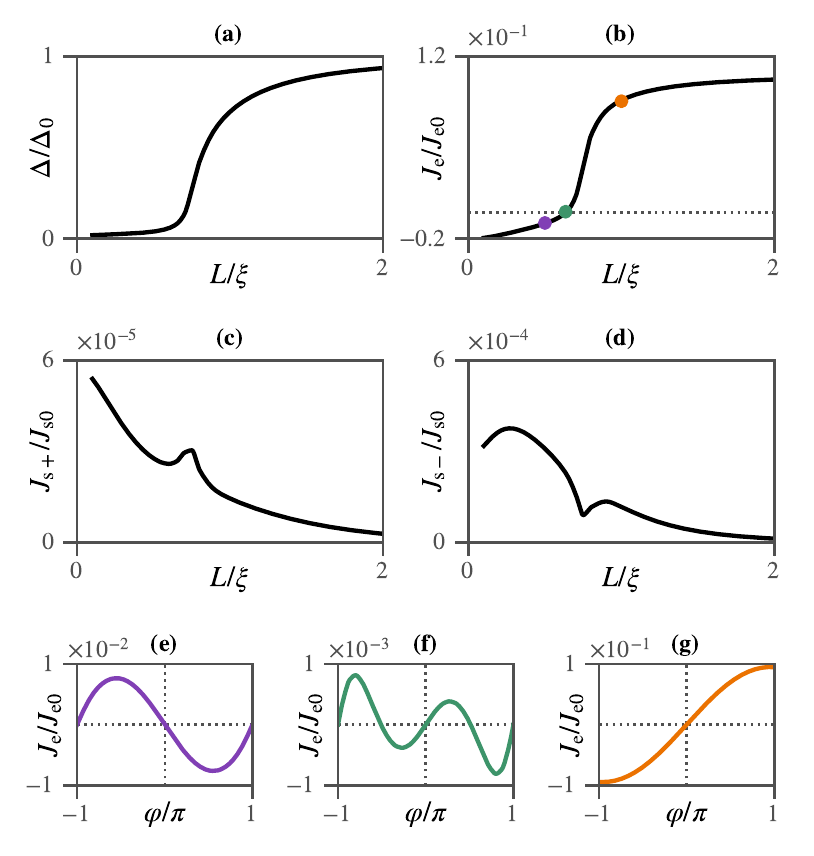}
    \caption
    {
      Plots of the (a) superconducting gap, (b) charge current, (c) exchange current, and (d) polarization current as functions of the superconductor length~$L$.
      The shape of the charge current-phase relation is shown below for a superconductor length (e) $L/\xi=0.500$, (f) $L/\xi=0.635$, and (g) $L/\xi=1.000$.
      The currents were calculated for  $\alpha_\text{sf} = \alpha_\text{so} = 0$, and a phase-difference $\varphi=\pi/2$ between source and drain.
      The lengths in (e--{g}) are indicated by coloured markers in (b).
    }
    \label{fig:length}
\end{figure}

\begin{figure}[tb!]
    \centering
    \includegraphics{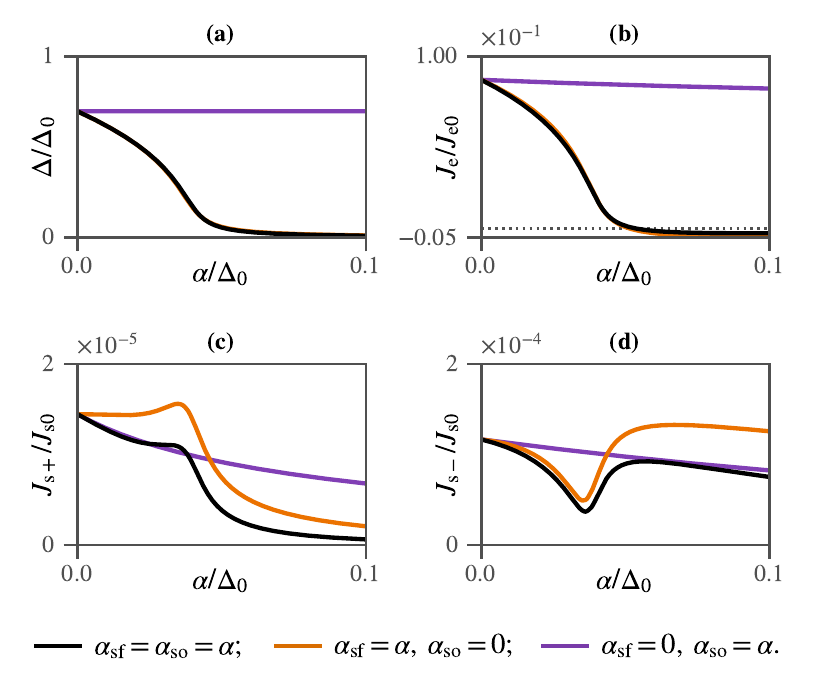}
    \vspace{-2ex}
    \caption
    {
      Plots of the (a) superconducting gap, (b) charge current, (c) exchange current and (d) polarization current as functions of the scattering rate $\alpha$.
      The black curves include both spin-flip and spin-orbit scattering; the orange curves show pure spin-flip scattering; the purple curves show pure spin-orbit scattering.
      Note that in (a) and (b), the black and orange curves overlap almost completely.
    }
    \label{fig:scattering}
\end{figure}

In \cref{fig:position}, we see how the charge and spin currents vary as functions of position in a junction with scattering rates $\alpha_\text{sf}=\alpha_\text{so}=0.01\Delta_0$ and a phase-difference~$\varphi = \pi/2$.
For comparison, these scattering rates~$\alpha$ are related to the scattering lengths~$\ell$ by $\alpha = \Delta_0 \xi^2/8\ell^2$.
Using the spin-flip lengths reported in \refcite{BassPratt2007}, we find that $\alpha_\text{sf}/\Delta_0$ varies in the range \text{$10^{-4}$--$10^{-1}$} for typical nonmagnetic metals at cryogenic temperatures, but it can be further increased by doping with magnetic atoms.
We see that the charge current is preserved as expected, and since $J_\mathrm{s-}$ is interpreted as the spin-polarization of the charge current, this must necessarily also be dissipationless.
To consider the exchange current, we note that the spin current is carried by triplet pairs, and randomization by scattering merely alters the number of singlet/triplet states available, via the introduction of an imaginary energy term.
Since the impurities provide no means of conversion from triplet to singlet states, or rotation to another triplet state, the magnitude of the conserved current is governed by the size of the junction {(}\cref{fig:length}{)} and the scattering strength {(}\cref{fig:scattering}{)}, as discussed below.

The effect of the superconductor length is investigated in \cref{fig:length} for the ideal case without spin-dependent scattering.
From \cref{fig:length}(a), we see that as the length is increased, the superconducting gap~$\Delta$ increases from zero to the bulk gap~$\Delta_0$.
During this transition from a normal to superconducting state, the charge current~(b) switches sign, indicating that we have a \text{0--$\pi$} transition in the junction.
We also note that the charge current is significantly {increased} when the central layer is in a superconducting {rather} than normal state, as one should expect.
Investigating the current-phase relation {around} the \text{0--$\pi$} transition {(between~(e) and~(g) as indicated in~(b))}, reveals that the junction simultaneously switches from a $\sin(\varphi)$ to {$\sin(\varphi/2)\,\text{sgn}[\cos(\varphi/2)]$} {current-phase relation} during the transition~\cite{OuassouSinphin,Zapata1996,Ishikawa2001,Yu1999,DeLuca2009,Beenakker2013}.
Exactly at the 0--$\pi$ transition~(f), the current-phase relation is dominated by a higher-order $\sin(2\varphi)$ contribution.

As the superconductor length increases, the exchange~(c) and polarization~(d) currents decay.
This can be explained as follows.
The concentration of triplet pairs decays exponentially away from the interfaces, meaning that the triplet concentration at the center of the junction decreases exponentially with the superconductor length.
Since the spin currents are conserved, the triplet concentration at the center acts as a ``bottleneck'' for the spin currents.
Thus, the spin currents also decay exponentially with the system length.
One notable exception to this monotonic decay occurs precisely at the 0--$\pi$ transition point, where the spin currents suddenly fluctuate.
This happens because the charge current drops abruptly near the transition point, which causes the polarization current to decrease while the exchange current increases.

In \cref{fig:scattering}(a), we see how the scattering rates affect the gap.
Since singlet pairs are resistant to spin-orbit scattering but destroyed by spin-flip scattering, the gap is only significantly suppressed by the spin-flip impurities.
The majority of the charge current in a superconductor is of course transported by the condensate of singlet pairs, so the charge current (b) is also hindered by spin-flip but not spin-orbit scattering.
Just like we found in \cref{fig:length}, the system undergoes a 0--$\pi$ transition when the superconducting gap becomes nonzero, and the charge current-phase relation (not shown) changes from a $\sin(\varphi)$ to {$\sin(\varphi/2)\,\text{sgn}[\cos(\varphi/2)]$} shape during this phase-transition.

Looking at the spin currents (c--d), we note that an increasing spin-orbit scattering causes a weak monotonic decrease.
This is because the scattering suppresses the triplet population throughout the junction, limiting the size of the spin currents.
In the case of spin-flip scattering, however, the picture becomes more complex.
Small increases in the scattering rate decreases the polarization current -- but further increasing it leads to a resurgence.
In fact, it reaches a value that is even higher than for zero spin-flip scattering, leading to the counter-intuitive conclusion that a moderate spin-flip scattering can actually \emph{increase} the spin current.
For the exchange current, we see the opposite trend: it initially increases with the spin-flip scattering, but then decays rapidly afterwards.

{
  The reason for this behaviour is as follows.
  There are two triplet sources for the central superconductor: (i) singlet pairs from the central superconductor that are converted to triplets at the Co-interface; and (ii) singlet pairs from the outer superconductors that are converted to triplets as they diffuse through the Ho and Co layers.
  The first kind of triplet dominates the transport when the central gap is strong (${\Delta \rightarrow \Delta_0}$), but only the second kind contributes when it is weak (${\Delta \rightarrow 0}$).
  It can be shown that for our junction parameters, the first kind results in a positive contribution to the charge current and the second a negative, explaining the 0--$\pi$ transition as a function of any parameter that modulates the gap.
  The {$0$--$\pi$} transition in the triplet current occurs slightly earlier than the corresponding transition in the total charge current, but the transition regime matches the non-monotonic regimes in \cref{fig:scattering}(c--d) very well.
  Furthermore, the first kind of triplet is less spin-polarized than the second kind since it did not have to pass \emph{through} the Co spin-filters, explaining the increase of the polarization current in \cref{fig:scattering}(d) as the dominant triplet source changes.
}

\section{Discussion}
We have shown analytically and numerically that spin supercurrents are conserved in superconductors, even when we include spin-flip and spin-orbit scattering processes.
We proceeded to analyze how the results varied with sample size and scattering rates in \cref{fig:length,fig:scattering}, and identified a combination of a 0--$\pi$ transition, {$\sin(\varphi) \rightarrow \sin(\varphi/2)\,\text{sgn}[\cos(\varphi/2)]$} transition, and a complex modulation of the spin current.
Increasing the interface transparencies would have increased all currents in the junction, and choosing a material with higher polarization than Co should enhance the polarization current.

The conservation of spin currents can intuitively be understood as follows.
The polarization current {physically corresponds to} the spin-polarized part of the charge current.
Since Cooper pairs that participate in charge transport are protected against resistive scattering, the same protection applies to the polarization current as well, resulting in it being conserved.
The exchange current, on the other hand, can be shown to be conserved even in non-superconducting F/N/F systems~\cite{Chen2014}.
If both the polarization and exchange currents are conserved separately, the total spin current is conserved as well.

{It was noted in \refcite{Linder2009} that an equilibrium spin current is conserved in a superconductor free from magnetic impurities.
The conservation of the spin current is consistent with the angular momentum conservation law since spatial variations in the current must generate a torque on a magnetic order parameter, and no such torque is present in a superconductor.} 
{However, this argument breaks down in the presence of magnetic impurities: in this case, the spatial variations in the current could generate a torque on the impurity spin orientations.
Since we still find that the spin supercurrent is conserved in the presence of such impurities, the spin supercurrents must behave in a fundamentally different way from their dissipative counterparts.}

{As shown in \cref{fig:length}(c--d), the spin supercurrents are small if the central superconductor becomes much larger than the coherence length, which in our proposed junction is $\xi \approx 30$~nm.
Since the spin relaxation length can exceed 0.5~$\othermu$m in e.g. Al, the conservation of spin supercurrents over $\sim\!30$~nm may not seem that significant.
However, we {expect the spin supercurrent to have a conserved component also in ballistic junctions}.
Since the superconducting coherence length in e.g. Al can reach 1.6~$\othermu$m in the ballistic case, this entails a conservation over remarkable length scales.
Moreover, spin supercurrents are fundamentally different from conventional spin currents since they may be manipulated using phase-coherent circuits in equilibrium, which may lead to entirely new kinds of spintronic device design.
While we do not propose any device that supersedes its conventional equivalents here, {a thorough understanding} of the properties of spin supercurrents will be vital for the future development of such devices.}

Our numerical results show that the charge currents {are typically below} $10^{-1} J_{\mathrm{e0}} \approx 1$~MA/cm$^2$, which is reasonable for a supercurrent inside a superconductor. 
However, the spin currents are typically of order $10^{-4} J_{\mathrm{s0}} \approx {(\hbar/2e) \,\times\,} 1$~kA/cm$^2$, which is small compared to what is routinely produced in non-superconducting spintronics circuits.\footnote{For instance, charge current densities up to 120~MA/cm$^2$ have been achieved in permalloy \cite{PyCurrent}, which has a spin polarization of 0.38 \cite{Villamor2013}, thus yielding spin current densities of $(\hbar/2e)\times46$~MA/cm$^2$.}
The purpose of the junction studied in this paper was however not to maximize the spin currents, but to investigate the fundamental physics and demonstrate that spin currents are conserved for a completely general structure with both exchange and polarization currents.
It would, however, be straight-forward to enhance the spin currents by a few orders of magnitude.
For instance, one could increase the tunneling conductance towards $G_0$, reduce the thickness of the Ho layers to $\zeta/2$, and remove the Co spin filters.
The spin current is conserved also in these more optimized junctions, which may be of more interest for applications.
Morover, in our structure we used magnetically inhomogeneous Ho layers to generate the $\ket{\up\up}$ and $\ket{\dn\dn}$ triplet components, which are long-ranged in the Co spin filters.
However, intrinsic spin-orbit coupling can also perform this function \cite{BT2013,BT2014,Jacobsen2015}.
In that case, homogeneous ferromagnets would be sufficient \cite{JKL2016}, and it could be instructive to include such systems {in} further investigations of current maximization.

A numerical treatment of a ballistic S/F/S/F/S system in equilibrium was considered in \refcite{Halterman_Alidoust_2016}.
{In such a system the current magnitude decreases with layer length because it decreases the ferromagnetic coupling.}
In our case, we have {demonstrated the surprising result that} spin currents are immune to magnetic and spin-orbit impurities, {inevitably present in real materials}, which rapidly destroy spin currents in non-superconducting systems.
Furthermore, we have shown the counter-intuitive result that the {magnitude of the spin currents} can actually increase with impurity concentration.

\vspace{4ex}
\section{Conclusion}%
Using a combination of both analytical arguments and numerical simulations, we have shown that spin supercurrents are conserved in superconductors, even in the presence of spin-dependent scattering processes.
Furthermore, we have shown that the charge and spin supercurrents have a nonmonotonic dependence on the various junction parameters, and that the current-phase relation also changes its shape as these are varied.
The result that spin supercurrents do not decay in superconductors has profound consequences with regard to potential applications based on spintronics, since it implies that information carried by the spin degree of freedom can be transmitted without loss or decoherence through a superconductor.


\begin{acknowledgments}
  We thank Morten Amundsen and Vetle Risinggård for useful discussions, and acknowledge support from the {Research Council of Norway grant numbers 216700 and 240806 and the Outstanding Academic Fellows programme at NTNU.
  We also acknowledge support from NTNU and the Research Council of Norway for funding via the Center of Excellence \emph{QuSpin}.}
\end{acknowledgments}


\begin{thebibliography}{100}
\bibitem{LinderRobinson2015}
J. Linder and J.W.A. Robinson.
\textit{Nat. Phys.} \textbf{11}, 307 (2015).

\bibitem{EschrigRev2011}

M.~Eschrig.
\textit{Physics Today} \textbf{64}, 43 (2011).


\bibitem{EschrigRev2015}
M. Eschrig.
\textit{Rep. Prog. Phys.} \textbf{78}, 10 (2015).

\bibitem{BlamireRobinson2014}

M.G.~Blamire and J.W.A.~Robinson.
\textit{J. Phys. Condens. Matter} \textbf{26}, 453201 (2014).


\bibitem{Yang2010}
H. Yang et al.
\textit{Nat. Mater.} \textbf{9}, 586 (2010).

\bibitem{Quay2013}
C.H.L. Quay et al.
\textit{Nat. Phys.} \textbf{9}, 84 (2013).

\bibitem{Wakamura2015}
T. Wakamura et al.
\textit{Nat. Mater.} \textbf{14}, 675 (2015).

\bibitem{Li2013}
B. Li et al.
\textit{Phys. Rev. Lett.} \textbf{110}, 097001 (2013).

\bibitem{BassPratt2007}
J. Bass and W.P. Pratt Jr.
\textit{J. Phys.: Condens. Matter.} \textbf{19}, 183201 (2007).

\bibitem{OtaniKimura2011}

Y. Otani and T. Kimura.
\textit{Phil. Trans. R. Soc. A} \textbf{369}, 3136 (2011).\\[-2ex]


\bibitem{JKL2016}
S.H. Jacobsen, I. Kulagina and J. Linder.
\textit{Sci. Rep.} \textbf{6}, 23926 (2016).

\bibitem{Josephson1962}

B. Josephson.
\textit{Phys. Lett.} \textbf{1}, 251 (1962).


\bibitem{Golubov2004}

A.A. Golubov, M.Yu. Kupriyanov and E. Il'ichev.
\textit{Rev. Mod. Phys.} \textbf{76}, 411 (2004).


\bibitem{Buzdin2005}

A.I. Buzdin.
\textit{Rev. Mod. Phys.} \textbf{77}, 935 (2005).


\bibitem{Ryazanov2001}

V. Ryazanov et al.
\textit{Phys. Rev. Lett.} \textbf{86}, 2427 (2001).


\bibitem{Szombati2016}

D.B. Szombati et al.
\textit{Nat. Phys.} \textbf{12}, 568 (2016).


\bibitem{Buzdin2008}

A.I. Buzdin. 
\textit{Phys. Rev. Lett.} \textbf{101}, 107005 (2008).


\bibitem{Reynoso2008}

A. Reynoso et al.
\textit{Phys. Rev. Lett.} \textbf{101}, 107001 (2008).


\bibitem{Zazunov2009}

A. Zazunov et al.
\textit{Phys. Rev. Lett.} \textbf{103}, 147004 (2009).


\bibitem{Tanaka2009}

Y. Tanaka, T. Yokoyama, and N. Nagaosa.
\textit{Phys. Rev. Lett.} \textbf{103}, 107002 (2009).




\bibitem{OuassouSinphin}
J.A. Ouassou and J. Linder.
arXiv:1612.03177.

\bibitem{Zapata1996}

I. Zapata and F. Sols. 
\textit{Phys. Rev. B} \textbf{53}, 6693 (1996).


\bibitem{Ishikawa2001}

H. Ishikawa, S. Kurihara and Y. Enomoto.
\textit{Physica C} \textbf{350}, 62 (2001).


\bibitem{Yu1999}

M.Yu. Kupriyanov et al. 
\textit{Physica C} \textbf{326}, 16 (1999).


\bibitem{DeLuca2009}

R. De Luca and F. Romeo.
\textit{Phys. Rev. B} \textbf{79}, 094516 (2009).


\bibitem{Beenakker2013}

C.W.J. Beenakker.
\textit{Annu. Rev. Conden. Matter Phys.} \textbf{4}, 113 (2013).

\bibitem{Morten2004}
J.P. Morten, A. Brataas and W. Belzig.
\textit{Phys. Rev. B} \textbf{70}, 212508 (2004).

\bibitem{Morten2005}
J.P. Morten, A. Brataas and W. Belzig.
\textit{Phys. Rev. B} \textbf{72}, 014510 (2005).


\bibitem{Robinson2010}
J.W.A. Robinson, J.D.S. Witt and M.G. Blamire.
\textit{Science} \textbf{329}, 59 (2010).

\bibitem{Alidoust2014}
M. Alidoust and K. Halterman.
\textit{Physical Review B} \textbf{89}, 195111 (2014).


\bibitem{SnDos}
C. Yu, J. Liu, H. Lu and J. Chen.
\textit{Solid State Commun.} \textbf{140}, 538 (2006).

\bibitem{Jacobsen2015}
S.H. Jacobsen, J.A. Ouassou and J. Linder.
\textit{Phys. Rev. B} \textbf{92}, 024510 (2015).

\bibitem{Koehler1966}
W. Koehler, J. Cable, M. Wilkinson and E. Wollan.
\textit{Phys. Rev. B} \textbf{151}, 414 (1966).

\bibitem{Chiodi2013}
F. Chiodi et al. 
\textit{Europhys. Lett.} \textbf{101}, 37002 (2013).

\bibitem{Usadel1970}
K. Usadel.
\textit{Phys. Rev. Lett.} \textbf{25}, 507 (1970).




\bibitem{Rammer1986}

J. Rammer.
\textit{Rev. Mod. Phys.} \textbf{58}, 323 (1986).


\bibitem{Chandrasekhar2004}

V. Chandrasekhar.
\textit{The Physics of Superconductors} \textbf{2}, 55 (2004). \\[-2ex]


\bibitem{Belzig1999}

W. Belzig et al.
\textit{Superlattices and microstructures} \textbf{25}, 1251 (1999).


\bibitem{Silaev2015}

M. Silaev et al.
\textit{Phys. Rev. Lett.} \textbf{114}, 167002 (2015).


\bibitem{Ouassou2016a}
J.A. Ouassou, A. Di Bernardo, J.W.A. Robinson and J. Linder.
\textit{Sci. Rep.} \textbf{6}, 29312 (2016).

%


\bibitem{Kupriyanov1988}
M.Y. Kupriyanov and V.F. Lukichev.
\textit{Soviet Physics JETP} \textbf{67}, 1163 (1988).

\bibitem{Eschrig2015}
M. Eschrig, A. Cottet, W. Belzig and J. Linder.
\textit{New J. Phys.} \textbf{17}, 83037 (2015).

\bibitem{Machon2013}

P. Machon, M. Eschrig and W. Belzig.
\textit{Phys. Rev. Lett.} \textbf{110}, 047002 (2013).


\bibitem{Cottet2007}

A. Cottet.
\textit{Phys. Rev. B} \textbf{76}, 224505 (2007).


\bibitem{Cottet2009}

A. Cottet et al.
\textit{Phys. Rev. B} \textbf{80}, 184511 (2009).


\bibitem{Cottet2011}

A. Cottet et al.
\textit{Phys. Rev. B} \textbf{83}, 184511 (2011).


\bibitem{Villamor2013}
E. Villamor, M. Isasa, L.E. Hueso and F. Casanova.
\textit{Phys. Rev. B} \textbf{88}, 184411 (2013).

\bibitem{Chen2014}
W. Chen, P. Horsch and D. Manske.
\textit{Phys. Rev. B} \textbf{89}, 064427 (2014). 

%
%

\bibitem{Linder2009}
{J. Linder, T. Yokoyama and A. Sudbø.
\textit{Phys. Rev. B} \textbf{79}, 224504 (2009).}

\bibitem{Halterman_Alidoust_2016}
K. Halterman and M. Alidoust.
\textit{Supercond. Sci. Technol.} \textbf{29}, 055007 (2016).

\bibitem{BT2013}
F.S. Bergeret and I.V. Tokatly.
\textit{Phys. Rev. Lett.} \textbf{110}, 117003 (2013).

\bibitem{BT2014}
F.S. Bergeret and I.V. Tokatly.
\textit{Phys. Rev. B} \textbf{89}, 134517 (2014).

\bibitem{PyCurrent}
A. Yamaguchi et al.
\textit{Phys. Rev. Lett.} \textbf{92}, 077205 (2004).



\end{thebibliography}
\end{document}